# RESEARCH AND DEVELOPMENT WORKSTATION ENVIRONMENT: THE NEW CLASS OF CURRENT RESEARCH INFORMATION SYSTEMS

*O.V. Palagin, V.Yu. Velychko, K.S. Malakhov, O.S. Shchurov*

Against the backdrop of the development of modern technologies in the field of scientific research, the new class of Current Research Information Systems (CRIS) and related intelligent information technologies have arisen. It was called – Research and Development Workstation Environment (RDWE) – the comprehensive problem-oriented information systems for scientific research and development lifecycle support. The given paper describes design and development fundamentals of the RDWE class systems. The general information model of the RDWE class systems is developed. Also the paper represents the information model of the RDWE class system for supporting research in the field of ontology engineering – the automated building of applied ontology in an arbitrary domain area, scientific and technical creativity – the automated preparation of application documents for patenting inventions in Ukraine. It was called – Personal Research Information System. The main results of our work are focused on enhancing the effectiveness of the scientist's research and development lifecycle in the arbitrary domain area.

На фоні розвитку сучасних технологій в сфері наукових досліджень, виник новий клас засобів комп'ютерних систем і відповідних інтелектуальних інформаційних технологій, що підтримують основні етапи життєвого циклу наукових досліджень. Цей клас систем отримав назву – Автоматизоване робоче місце наукових досліджень (АРМ-НД) – складні проблемно-орієнтовані інформаційні системи підтримки повного циклу наукових досліджень. В роботі наведено основи проектування і розробки систем класу АРМ-НД, розроблено узагальнену інформаційну модель систем класу АРМ-НД, а також наведена інформаційна модель розробленої АРМ-НД системи підтримки науково-технічної творчості та досліджень в області онтологічного інжинірингу. Отримані результати орієнтовані на підвищення ефективності повного науково-дослідного циклу роботи наукових співробітників в довільних предметних галузях.
Ключові слова: АРМ-НД; АРМ; хмарне середовище; онтологічний інжиніринг; композитний веб-сервіс; хмарні обчислення; хмарне середовище навчання.

На фоне развития современных технологий в сфере научных исследований, возник новый класс средств компьютерных систем и соответствующих интеллектуальных информационных технологий, поддерживающих основные этапы жизненного цикла научных исследований. Этот класс систем получил название – Автоматизированное рабочее место научных исследований (АРМ-НИ) – сложные проблемно-ориентированные информационные системы поддержки полного цикла научных исследований. В работе описаны основы проектирования и разработки систем класса АРМ-НИ, разработана обобщённая информационная модель систем класса АРМ-НИ, а также описана информационная модель разработанной АРМ-НИ системы поддержки научно-технического творчества и исследований в области онтологического инжиниринга. Полученные результаты ориентированы на повышение эффективности полного научно-исследовательского цикла работы научных сотрудников.
Ключевые слова: АРМ-НИ; АРМ; облачная среда; онтологический инжиниринг; композитный веб-сервис; облачные вычисления; облачная среда обучения.

## Introduction

The development of modern technologies increasingly covers the field of intellectual activity and, especially, in the field of scientific research and development. The new class of Current Research Information Systems and related intelligent information technologies have arisen that support the main stages of the scientific research and development lifecycle, starting with the semantic analysis of the information & data material of arbitrary domain area and ending with the formation of constructive features of innovative proposals. It was called – *Research and Development Workstation Environment* (RDWE) – the comprehensive problem-oriented information systems for scientific research and development support. A distinctive feature of such systems and technologies is the possibility of their problematic orientation to various types of scientific research and development by combining on a variety of functional services and adding new ones within the *Cloud-integrated Environment* (inside the Ubuntu open source operating system as the integrating cloud environment for instance).

## Current research information systems

In the modern English-speaking scientific environment, a steady term – *Current Research Information System* (CRIS) [1] was introduced to designate scientific information systems for access to scientific and academic information. It is important to emphasize that the definition of CRIS also specifies that CRIS is not only intended for direct access to information sources of science but also, according to the ERGO project [2], for:

- to facilitate access to national scientific and technical information services;
- to identify the main existing information sources and to evaluate access possibilities and the potential for the utilization of these sources at European level;
- to invite national data hosts to offer their Research and Development (R&D) information and to make this information searchable for the user.

The EuroCRIS [3] organization was founded in 2002, is an international not-for-profit association, that brings together experts on research information in general and research information systems. The mission of EuroCRIS is to promote cooperation within and share knowledge among the research information community and interoperability of research information through CERIF [4] – the Common European Research Information Format. Areas of interest also cover research databases, CRIS related data like scientific datasets, (open access) institutional repositories, as well as data access and exchange mechanisms, standards and guidelines and best practice for CRIS. The EuroCRIS provides the framework for the flow of information/data between a broad variety of stakeholders: researchers, research managers and administrators, research councils, research funders, entrepreneurs, and technology transfer organizations.

The basic user categories of modern CRIS and their information demands are listed in [5] and represented in table 1.

Table 1. The basic categories of CRIS and their information demands

| Users group | Information demands |
|---|---|
| Researchers | Search for scientific results, collaborators, equipment and material resources, projects financing. Easy access to relevant information and associated software, processor power, storage systems and detectors to collect more data to overcome incomplete or inconsistent information – the paradigm of *Service-Oriented Science (SOS)* in which web services are virtual access points to data, computational resources, R&D environments [6]. |
| Teachers, students | Implementation of the recent research results in the educational process and mastering of research results in the learning process. The organization of service support of e-learning at the university, especially at the *replaced university* [7]. Applying to the educational process the modern paradigm of *Service-Oriented Learning (SOL)* implemented as *Personal Learning Environments*, *Virtual Learning Environments, Learning Management Systems*, *Education-as-a-Service* models*, and *Cloud Learning Environments* – using RDWE systems as an E-learning/education-oriented environment. |
| Institute's board of administrators, analysts | Research management and administration, statistical reporting on recent achievements, determining the role of research institutes and researchers in the research process. |
| Experts, financiers, organizations that provide financial resources | Project reports, estimation of expected and obtained research results. |
| Research project chairman's | Research processes coordination, search for collaborators and search for research funders for project financing. |
| Investors, research funders, industry | Searching for technology, engineering, and hunting experts who can evaluate the technology. |
| Community and media | Acquisition information and data in the accessible form for perception and understanding. |

This user's groups and their information demands determine the main types of information/data sources that CRIS works with: progress reports, project results reports, personal information and data, scientific publication, organizations, projects, research results, technology and engineering, patents, fund programs, expert assessments, digital libraries, encyclopedias and dictionaries, websites, mailing lists, social networks, databases, computing resources, technological normative and other documents, educational and museum resources. Also, this user's groups and their information demands determine the main types of services provided by CRIS: reuse of scientific developments, methodologies, and technologies; information search; targeted dissemination of information; messages services; bridging of horizontal and vertical relations between organizations; backup data storage and archival information/data repository; support for the educational process; providing analytical services.

Requirements for CRIS in the information management aspect of strategic research management has been analyzed in [8], which describes the types of research managerial activities, introduces the currently available information sources and how the information found in these is applied today.

The paper [9] describes requirements for CRIS to effective dissemination of technologies, the management of scientific programs and functioning of funds for research funding. Noted the importance of CRIS systems to collaborate researchers and to support the information functioning of funds. Also represented the basic lifecycle of scientific programs and the information demands of participants at each phase of the cycle.

The CORDIS [10] portal emphasized the basic use cases of scientific portals for researchers:

- keep up-to-date on current research findings and strategic directions;
- identify funding sources for R&D;
- find partners to cooperate in R&D activities and share expertise;
- form transnational consortia for exploitation of research results;
- promote and locate transferable technologies, and more.

In the field of scientific communication and collaboration next tools are used: telecommunications services of direct interaction (video chat, video conference, voice calls, text and video messages) – Skype, Viber, VSee etc.; cloud-based sets of team collaboration tools and services – Slack etc.; web conference platforms for various types of online collaborative services including webinars, webcasts, and peer-level web meetings – Adobe Connect etc.; services of indirect communication – email; social services on the Internet – social networks with instant messaging such as Facebook and Telegram. These services are widespread with off-the-shelf software and in the availability of communication bus with sufficient bandwidth allow instant scientific communication.

CRIS can include systems that combine social network technologies as well as usual file repository technologies. As commercial projects, such CRIS is oriented to one of the following business models:

- to sell user's and subscriber's data to advertisers that provide targeted ads, which usually costs more than usual. Also, the platform offers universities and corporations the option to post job listings for free, but generates revenue by charging some money for increased visibility (ResearchGate) [11, 12];

- to carry out analytics for the uploaded content and recommend suitable research activity for a fee. Also, the platform generates revenue by providing trending research data to R&D institutions that can improve the quality of their decisions (Academia.edu) [12, 13];

- to provide additional paid services for storage of data (publications, raw data, and media files) – the individual researcher is served with a freemium model, users can upgrade their cloud space for a monthly fee; to organize online discussion community platforms, forums for sharing and discussing research. Also, the platform offers real-time analytics to libraries for a fee and has opened its application programming interface (API) for programmers who build third-party apps on top of the data generated across the platform (Mendeley) [12, 14].

According to demands of the scientific community members Academia.edu and ResearchGate provide the following basic features for registered users:

- members may "follow" a research interest, in addition to following other individual members;
- allows its users to create a profile, upload their work(s), select areas of interests;
- platform indexes self-published information on user profiles to suggest members connect with others who have similar interests.
- private chat rooms where users can share data, edit shared documents, or discuss confidential topics;
- a research-focused job board feature;
- ResearchGate publishes a citation impact measurement in the form of an "RG Score";
- Academia.edu proclaims it supports the open science or open access movements and instant distribution of research, and a peer-review system that occurs alongside distribution, instead of prior to it.

However, the Academia.edu service is not designed to provide comprehensive information support for scientific research, and users, in general, posts published articles to the system. Unlike Academia.edu, ResearchGate has more diverse features. People that wish to use the site need to have an email address at a recognized institution or to be manually confirmed as a published researcher to sign up for an account. Members of the site each have a user profile and can upload research output including papers, data, chapters, negative results, patents, research proposals, methods, presentations, and software source code. Users may also follow the activities of other users and engage in discussions with them. Users are also able to block interactions with other users. ResearchGate has a blogging feature for users to write short reviews on peer-reviewed articles. When a member posts a question, it is fielded to others that have identified on their user profile that they have a relevant expertise.

The social networking Internet services Academia.edu and ResearchGate are largely aimed at forming the so-called "name" of the scientist and serving a research-focused job board feature.

## General design principles (approaches) of the CRIS-like systems for scientific R&D support

There are different approaches to define the structure of research activities and the *research lifecycle* [15, 16]. The review of these approaches, as well as the analysis of research activities, allows defining its basic types, which regulate its structure and are invariant in relation to the domain area, content, methods, and approaches of the research activity. The basic types of research activities are [16]:

- *Descriptive* vs. *Analytical*: Descriptive research includes surveys and fact-finding enquiries of different kinds. The major purpose of descriptive research is the description of the state of affairs as it exists at present. In social science and business research, we quite often use the term *Ex post facto research* for descriptive research studies. The main characteristic of this method is that the researcher has no control over the variables; he can only report what has happened or what is happening. Most ex-post facto research projects are used for descriptive studies in which the researcher seeks to measure such items as, for example, the frequency of shopping, preferences of people, or similar data. Ex post facto studies also include attempts by researchers to discover causes even when they cannot control the variables. The methods of research utilized in descriptive research are survey methods of all kinds, including comparative and correlational methods. In analytical research, on the other hand, the researcher must use facts or information already available and analyze these to make a critical evaluation of the material;

- *Applied* vs. *Fundamental*: Research can either be applied (or action) research or fundamental (to basic or pure) research. Applied research aims at finding a solution for an immediate problem facing a society or an industrial/business organization, whereas fundamental research is mainly concerned with generalizations and with the formulation of a theory. Research concerning some natural phenomenon or relating to pure mathematics are examples of fundamental research. Similarly, research studies, concerning human behavior carried on with a view to generalize about human behavior, are also examples of fundamental research, but research aimed at certain conclusions (say, a solution) facing a concrete social or business problem is an example of applied research. Research to identify social, economic, or political trends that may affect an institution or the copy research (research to find out whether certain communications will be read and understood) or the marketing research or evaluation research are examples of applied research. Thus, the central aim of the applied research is to discover a solution for some pressing practical problem, whereas basic research is directed towards finding information that has a broad base of applications and thus, adds to the already existing organized body of scientific knowledge;

- *Quantitative* vs. *Qualitative*: Quantitative research is based on the measurement of quantity or amount. It is applicable to phenomena that can be expressed in terms of quantity. Qualitative research, on the other hand, is concerned with the qualitative phenomenon, i.e., phenomena relating to or involving quality or kind. For instance, when we are interested in investigating the reasons for human behavior (i.e., why people think or do certain things), we quite often talk of 'Motivation Research', an important type of qualitative research. This type of research aims at discovering the underlying motives and desires, using in-depth interviews for the purpose. Other techniques of such research are word association tests, sentence completion tests, story completion tests and similar other projective techniques. Attitude or opinion research i.e., research designed to find out how people feel or what they think about a subject or institution is also qualitative research. Qualitative research is especially important in the behavioral sciences where the aim is to discover the underlying motives of human behavior. Through such research, we can analyze the various factors which motivate people to behave in a manner or which make people like or dislike a thing. It may be stated, however, that to apply qualitative research in practice is relatively a difficult job and therefore, while doing such research, one should seek guidance from experimental psychologists;

- *Conceptual* vs. *Empirical*: Conceptual research is that related to some abstract idea(s) or theory. It is generally used by philosophers and thinkers to develop new concepts or to reinterpret existing ones. On the other hand, empirical research relies on experience or observation alone, often without due regard for system and theory. It is data-based research, coming up with conclusions which are capable of being verified by observation or experiment. We can also call it as the experimental type of research. In such a research, it is necessary to get at facts firsthand, at their source, and actively to go about doing certain things to stimulate the production of desired information. In such a research, the researcher must first provide himself with a working hypothesis or guess as to the probable results. He then works to get enough facts (data) to prove or disprove his hypothesis. He then sets up experimental designs which he thinks will manipulate the persons or the materials concerned so as to bring forth the desired information. Such research is thus characterized by the experimenter's control over the variables under study and his deliberate manipulation of one of them to study its effects. Empirical research is appropriate when the proof is sought that certain variables affect other variables in some way. Evidence gathered through experiments or empirical studies is today considered to be the most powerful support possible for a given hypothesis;

- *Some Other Types of Research*: All other types of research are variations of one or more of the above-stated approaches, based on either the purpose of research, or the time required to accomplish research, on the environment in which research is done, or based on some other similar factor. Form the point of view of time, we can think of research either as *one-time research or longitudinal research*. In the former case, the research is confined to a single time-period, whereas in the latter case the research is carried on over several time-periods. Research can be *field-setting research or laboratory research or simulation research*, depending upon the environment in which it is to be carried out. Research can as well be understood as *clinical or diagnostic research*. Such research follows case-study methods or in-depth approaches to reach the basic causal relations. Such studies usually go deep into the causes of things or events that interest us, using very small samples and very deep probing data gathering devices. The research may be *exploratory* or it may be formalized. The objective of exploratory research is the development of hypotheses rather than their testing, whereas formalized research studies are those with substantial structure and with specific hypotheses to be tested. *Historical research* is that which utilizes historical sources like documents, remains, etc. to study events or ideas of the past, including the philosophy of persons and groups at any remote point of time. Research can also be classified as conclusion-oriented and decision-oriented. While doing *conclusion-oriented research*, a researcher is free to pick up a problem, redesign the enquiry as he proceeds and is prepared to conceptualize as he wishes. Decision-oriented research is always for the need of a decision maker and the researcher, in this case, is not free to embark upon research according to his own inclination. Operations research is an example of decision-oriented research since it is a scientific method of providing executive departments with a quantitative basis for decisions regarding operations under their control.

The general principle that should be considered when designing CRIS-like systems for scientific research and technical research creativity support is that the system must fully cover all the basic types of research activities, implement the full research lifecycle. Based on this basic principle can identify the main entities of CRIS-like systems [17]:

1. User with all related attributes.
2. Organizations which are related to users.
3. Organizations and foundations (research funders etc.) that provide grant support for scientific research.
4. Scientific communities with the ability to carry on a discussion between united users (analogues of chats and forums).
5. Scientific journals and publishers.
6. Bibliographies that can be created both by the users themselves and by the communities.
7. Publications of users that can be linked to publishers and periodicals.
8. Projects that can integrate users, bibliographies, documents, and general data, and can also be related to grants of research funders and organizations.

It should be noted that all these and other entities that may appear in the future should be able to communicate with each other. This communication will allow the flexible and comprehensive solution of the tasks of organizing of co-operative research activities by communities of scientists, to receive various reports, for example, lists of publication, lists of publisher's journals, statistical reports on the research activity of employees, etc. The integrated system should be a universal instrument or utility toolkit not only for the purposes of supporting scientific research but also for the analysis of the activities of scientific organizations, foundations, and publishers.

Another important principle for developing the CRIS-like system is the choice of a user identification model. There is no need to create multiple user accounts in different network services. For identification, it is suitable to choose one of the open identifying system: *Open Researcher and Contributor ID* (ORCID) [18], ResearcherID [19] or OpenID [20]. For scientists, it is suitable to choose ORCID – is a nonproprietary alphanumeric code to uniquely identify scientific and other academic authors and contributors. It is implemented in the open project ORCID. The ORCID organization offers an open and independent registry intended to be the de facto standard for contributor identification in research and academic publishing. The ORCID community includes everyone in the research community who recognizes the need for researchers to be uniquely identified and connected with their contributions and affiliations. This means individual researchers, organizations involved in research – universities, laboratories, commercial research companies, research funders, publishers, patent offices, data repositories, professional societies, and more – as well as organizations that build systems that support information management among and between these groups. Third-party tools allow the migration of content from other services into ORCID, for example, Mendeley2ORCID, for Mendeley.

## General requirements to the CRIS-like systems for scientific R&D support

In [4, 17, 21] reviewed general requirements and specifications for the CRIS-like systems. Summing up all these papers and resources, we can point out the most general requirements and specifications for the CRIS-like systems.

1. *Data coverage.* Coverage of the maximum amount of information resources and repositories. It is necessary to create data input procedures and data entry points to collect information. The following options for inputting data are available:
   - interactive user input;
   - data collection over the Internet uses the specialized software – web crawlers, Internet bots, intelligent agents etc.;
   - data exchange and CRIS integration with other CRIS-like systems.

2. *Data relevance.* The automatic collection of information over the Internet may accumulate out of date data. This problem can be solved by:
   - to create detailed formats for submitting metadata about resources, directories for thematic classification of resources, as well as adding metadata to resources. It is problematic to require authors clearly and accurately comply with the metadata formats and carefully add metadata to text sources that require additional work and explore metadata formats;
   - to classify and categorize all the information resources collected by experts (users) and web crawlers. Also, to indicate the degree of probability of belonging to the search query depending on its source;

▪ the exact indication of the search area to the search engines, as well as the criteria for the quality of the collected information;

▪ to create the ontologies of resources according to the user demands. The classification of resources by experts in accordance with these ontologies.

3. *Completeness and validity of data.* The data completeness problem is solved by the same method as the coverage data problem solved. The validity of data problem can be solved by:

▪ for interactive input – the restriction of input only by authenticated users;

▪ for the automated search engines – the restriction activity area of the web crawler to collect information;

▪ for input data exchange with other systems – by establishing precise filters for imported information resources;

▪ for all systems – verification and classification of the input data.

4. *The maintenance of the intelligent user query service.* User query services should support search by attributes, resource overviews by category, and full-text search using semantic document analysis.

5. *Support various levels of abstraction for data representation.* Internet efficiency for CRIS-like systems is not a direct consequence of the amount of available information/data or even its quality but is a direct consequence of the rate and accuracy of the selection of information for the demands of researchers. Support various levels of abstraction for information/data representation allows precipitating the information/data search without its loss. CRIS-like systems need to support several levels of abstraction, and not just full-text descriptions. It is necessary to represent information in a compact form (document ontology, abstract, annotation, compendium, or recap) selectively – from brief descriptions for the maximum quick search to very detailed descriptions of information objects.

6. *The historicity of the scientific information.* The specificity of the scientific information is quite a short period of its relevance. For many types of the scientific information resources, it is important to keep the description of the lifecycle of these resources and be able to restore the status of the resource at any given time.

7. *Maintaining an archival repository of scientific information.* As it was noted above, the most part of the scientific information fast becomes outdated. But there are information resources that may be relevant for a long time. These include: the long-term documents, the patents, and the multimedia information about historical events. Scientific reports of institutes, speeches of scientists and researchers can also have a huge historical value, becoming later even more valuable. Therefore, the system should support the possibility of long-term storage of the information resources.

8. *Distributed architectures support for the information systems.* This requirement is a prerequisite for completeness, authenticity, and relevance of information. The experience of usage of the CRIS-like systems has shown that it is difficult to implement, and in many cases, even impossible, the creation of centralized scientific systems that covers scientific information in some domain area of science, or in some country. In [22] Hale University researchers have proven proposals for the creation of administrative and technical mechanisms for the functioning of virtual distributed scientific libraries. In terms of the distributed computing environment, CRIS-like systems meet the following requirements:

▪ information exchange protocols support with other information systems;

▪ accepted metadata standards support for the export/import of data and the usage of metadata to describe and to specify resources;

▪ information verification feature;

▪ linking to internal resources support for both: in user interfaces and at the system level;

▪ online service for accessing links to other websites and documents;

▪ selection of resources in the intellectual process, according to published quantitative and thematic criteria;

▪ availability of descriptions of the resource's content in the range from brief annotation to detailed review;

▪ intelligently created structure or scheme for navigation in the resource space. A good example of such structure can be an ontological scheme of resources.

For some domain areas of research, it is also important that the scientific information system provides the ability to use computational resources that are based on generally accepted standards.

The basic concepts of the *workstation environment* (WE) consider to [23, 24] are – is an individual complex of information resources, software, hardware, organizational and technological facilities (tools) for individual or collective usage, united to perform certain functions of the employee. WE provide the worker with all the tools and facilities necessary to perform certain functions.

• to solve a certain class of tasks and problems, united by a common information processing technology;

• to formalize professional knowledge – the possibility of providing with the help of WE the automation of new functions and solving new tasks in the process of accumulation of experience with the system;

• to automate data processing in a specific area of activity;

- to provide the ability of data processing by the user itself;
- to provide user access to the set of software; hardware and information tools & facilities;
- to create custom comfortable working conditions and friendly user interface to communicate & interact with the system;
- to combine the system with other elements of information processing and modification, expansion capabilities of WE without interrupting its operation.

The structure and composition of the elements of any WE depend on its purpose, the composition of tasks, structure, software, the method of fixing data in primary documents and more. The WE structure is composed of:

− the functional part of WE;

− the software part of WE;

− the hardware part of WE;

− the information support part of WE. This is a set of facilities and methods for building an information base, which is divided into a machine and internal machine.

## General principles of the Research and Development Workstation Environment design

Scientific and research communities create networks that combine digital libraries, file repositories, archival information/data repositories, web servers with scientifically relevant information. The key objective factor that must be taken into account when designing, developing and using the workstation for the Ukrainian researcher is the limited financial resources appropriated for the development and maintenance of the workplace software, which results in the following consequences: the inability to involve the necessary number of specialists to the development process; minimizing the costs of the software implementing, maintaining and updating; system development for several years by the group of developers whose composition (body) is changing; the necessity to maximize the reuse of software that has already been developed. Next we define the general development principles of the RDWE and its components.

1. *Modular design.* The development of any information system, especially that aimed at solving the complex variety of problems and tasks, should be based on a modular design principle [17]. Modular design, or "modularity in design", is a design approach that subdivides a system into smaller parts called modules or skids, that can be independently created and then used in different systems. A modular system can be characterized by functional partitioning into discrete scalable, reusable modules; rigorous use of well-defined modular interfaces; and making use of industry standards for interfaces. When creating a modular system, instead of creating a monolithic application (where the smallest component is the whole), several smaller modules are written separately so that, when composed together, they construct the executable application program. Typically, these are also compiled separately, via separate compilation, and then linked by a linker. A just-in-time compiler may perform some of this construction "on-the-fly" at runtime. This makes modular designed systems if built correctly, far more reusable than a traditional monolithic design, since all (or many) of these modules may then be reused (without change) in other projects.

2. *Microservices-based architecture (MSA).* Microservices is a variant of the service-oriented architecture architectural style that structures an application as a collection of loosely coupled services [25]. In MSA, services should be fine-grained and the protocols should be lightweight. The benefit of decomposing an application into different smaller services is that it improves modularity and makes the application easier to understand, develop and test. There is no industry consensus yet regarding the properties of microservices, and an official definition is missing as well. Some of the defining characteristics that are frequently cited include [26]: services in MSA are often processes that communicate with each other over a network in order to fulfill a goal using technology-agnostic protocols such as HTTP; services in MSA should be independently deployable; services are easy to replace; services are organized around capabilities, e.g., user interface front-end, recommendation, logistics, billing, etc.; services can be implemented using different programming languages, databases, hardware and software environment, depending on what fits best; services are small in size, messaging enabled, bounded by contexts, autonomously developed, independently deployable, decentralized and built and released with automated processes.

3. *Cross-platform software (also multi-platform software or platform-independent software).* The researcher must have access to the services provided by the RDWE from various types of devices operating under the control of various operating systems. One of the ways to achieve this goal can be access to services through an adaptive user interface of the web applications. Web applications are typically described as cross-platform because, ideally, they are accessible from any of various web browsers within different operating systems. Such applications generally employ a client-server system architecture and vary widely in complexity and functionality. This wide variability significantly complicates the goal of cross-platform capability, which is routinely at odds with the goal of advanced functionality. Basic web applications perform all or most processing from a stateless server and pass the result to the client web browser. All user interaction with the application consists of simple exchanges of data requests and server responses. Implementation mechanisms:

- Node.js [27] server platform – is an open-source, cross-platform JavaScript run-time environment for executing JavaScript code server-side. Node.js enables JavaScript to be used for server-side scripting and runs scripts server-side to produce dynamic web page content before the page is sent to the user's web browser. Consequently, Node.js has become one of the foundational elements of the "JavaScript everywhere" paradigm, allowing web application development to unify around a single programming language, rather than rely on a different language for writing server-side scripts. Node.js brings event-driven programming to web servers, enabling development of fast web servers in JavaScript. Developers can create highly scalable servers without using threading, by using a simplified model of event-driven programming that uses callbacks to signal the completion of a task. Node.js connects the ease of a scripting language (JavaScript) with the power of Unix network programming. Node.js applications can run on most modern operating systems (Linux, macOS, Microsoft Windows and Unix servers) and hardware platforms (x86, x64, ARMv6, ARMv7, ARMv8).

- The adaptive user interface of the web application with responsive web design using Pug (is a template engine for Express), Bootstrap (is a design and style framework), JQuery library and AJAX (an approach to building user interfaces for web applications) – is a web design approach aimed at crafting the visual layout of sites to provide an optimal viewing experience – easy reading and navigation with a minimum of resizing, panning, and scrolling – across a wide range of devices, from mobile phones to desktop computer monitors.

- Storage service powered by MongoDB – is a free and open-source cross-platform document-oriented database program. Classified as a NoSQL database program, MongoDB uses JSON-like documents with schemas.

4. *JSON/XML data model and REST architectural style (RESTful web services)*. JavaScript Object Notation or JSON – is an open-standard file format that uses human-readable text to transmit data objects consisting of attribute-value pairs and array data types (or any other serializable value). It is a very common data format used for asynchronous browser-server communication. JSON standard defines a metalanguage, based on which, by imposing restrictions on the structure and content of documents, specific, domain-oriented markup languages are determined. The author of the document creates its structure, builds the necessary links between the elements, and uses those commands that meet its requirements and asks for the type of markup that it needs to handle the documents. Creating the correct structure of the information exchange mechanism at the beginning of the system project development, many future problems can be avoided due to the incompatibility of the data formats that are used by the various components of the system. *Representational state transfer* (REST) [28] or RESTful web services are a way of providing interoperability between computer systems on the Internet. REST-compliant Web services allow requesting systems to access and manipulate textual representations of Web resources using a uniform and predefined set of stateless operations. In a RESTful Web service, requests made to a resource's URI will elicit a response that may be in XML, HTML, JSON or some other defined format. The response may confirm that some alteration has been made to the stored resource, and it may provide hypertext links to other related resources or collections of resources. Using HTTP, as is most common, the kind of operations available include those predefined by the HTTP methods GET, POST, PUT, DELETE and so on. The RESTful architectural style possesses the following constraints [29]:

- *Client/Server*: Separation of concerns, exemplified by a client–server architecture. The idea is that different components can evolve independently—the user interface in the client can evolve separately from the server, and the server is simpler.

- *Stateless*: The client–server interaction is stateless. There is no stored context on the server. Any session information must be kept by the client.

- *Cacheable*: Data in a response (a response to a previous request) is labeled as cacheable or non-cacheable. If it is cacheable, the client (or an intermediary) may reuse that for the same kind of request in the future.

- *Uniform Interface*: There is a uniform interface between components. In practice, there are four interface constraints: resource identification – requests identify the resources they are operating on (by a URI, for example); resource manipulation through the representation of the resource – when a client or server that has access to a resource, it has enough information based on understanding the representation of the resource to be able to modify that resource; messages are self-descriptive – the message contains enough information to allow a client or server to handle the message, this is normally done through the use of Internet Media types (MIME types); use of hypermedia to change the state of the application – for example, the server provides hyperlinks that the client uses to make state transitions.

- *Layered System*: Components are organized in hierarchical layers; the components are only aware of the layer within which the interaction is occurring. Thus, a client connecting to a server is not aware of any intermediate connections. Intermediate filter components can change the message while it is in transit: because the message is self-descriptive and the semantics are visible, the filter components understand enough about the message to modify it.

- *Code on Demand*: Code on demand is optionally supported, that is, clients can download scripts that extend their functionality.

5. *Free/Libre and open-source software (FLOSS)*. FLOSS is software that can be classified as both free software and open-source software. That is, anyone is freely licensed to use, copy, study, and change the software in any way, and the source code is openly shared so that people are encouraged to voluntarily improve the design of the software. The benefits of using FLOSS can include decreased software costs, increased security, and stability (especially regarding malware), protecting privacy, and giving users more control over their own hardware. The use of free software will significantly save money at the stages of development and implementation of the RDWE.

6. *Cloud computing*. Cloud computing [30] can provide interaction between different CRIS-like systems (such as RDWE) over the Internet, with the optimal distribution of load between local and remote servers. In cloud computing paradigm, computer resources and capacity available to the user as Internet services for data processing. Cloud-computing providers offer their "Services" according to different models, of which the three standard models per NIST (The National Institute of Standards and Technology) [31] are *Infrastructure as a Service* (IaaS), *Platform as a Service* (PaaS), and *Software as a Service* (SaaS). These models offer increasing abstraction; they are thus often portrayed as a layer in a stack: infrastructure-, platform- and software-as-a-service, but these need not be related.

7. *Data storage based on the abstraction of data repository*. Abstraction of software modules from the implementation of data storage technology allows the system administrator to choose the type of data repository, which is maximally adapted for the purposes of the specific RDWE configuration.

8. *Data repositories synchronization*. The ability of the off-line operations of the RDWE and data synchronization with the central data repository. This is important in the absence of a permanent network connection. This feature implemented with the internal mechanisms for replication of the data repository.

## General information model of the Research and Development Workstation Environment

The RDWE class system's generalized information model represented as a three-tuple *composite web service* (CWS) using the revised formalism given in [32]:

$$CWS = \langle AWS, F, CI_{Env} \rangle,$$

where:

$CWS$ is the RDWE composite web service;

$AWS = \{aws_i | i = \overline{1,m}\}_{m \in N}$ is a set of atomic web services (problem-oriented microservices and FLOSS applications; personalized FLOSS applications) available for usage. The $AWS$ set consists of the problem-oriented atomic web services $aws$ and each of them can be designed and developed as a microservice or a desktop application, that allows them to be used as an independent software separately from the RDWE and as its components inside $CI_{Env}$;

$F = AWS : \{C_j | j = \overline{1,n}\}_{n \in N}$ is a set of functions, the functional filling-up of the RDWE, each function is the result of coordination and interaction of the $AWS$ elements;

$C_j \subseteq AWS, C_j = \{aws_k | k \geq 1, k \leq m\}_{k \in N}$ is a subset of atomic web services that are required to implement the $j$-th function of $CWS$;

$CI_{Env} = \{prl, mid, os, crd, typical_{FLOSS}\}$ is a set of elements (represented as layers) that combine into the Cloud-integrated Environment (CIE);

$prl$ – *physical resource layer* represents physical hardware and facility resources;

$mid$ – *middle layer* (using in the concepts of cloud service orchestration model [30]) represents resource abstraction and control layer. It is supposed to use OpenStack software platform;

$os$ – *operating system layer* represents guest operating system. It is supposed to use Ubuntu server with LXDE (abbreviation for Lightweight X11 Desktop Environment) desktop environment or Xfce desktop environment. Atomic web services work on the operating system layer – $aws_i \in os$;

$crd$ – *coordination component*. The $crd$ function is to coordinate atomic web services in $CWS$, and by the coordination procedure we will understand the execution of invocation of some $aws_k \in C_j$ in the defined sequence. The coordination component $crd$ can be implemented as the reverse proxy server of tasks. Nginx also is a part of $crd$ – used as front-end to control and protect access to the server on a private network, performs tasks such as load-balancing, authentication, decryption, and caching.

$typical_{FLOSS}$ – *a typical FLOSS layer* includes some regular application suit needed for the scientific research and development lifecycle (regular software suit may change in the future): LibreOffice office suite; Mozilla Firefox and Chromium web browsers; Sylpheed email client; Sublime Text 2 and jEdit source code editors; Wine compatibility layer that aims to allow computer programs developed for Microsoft Windows to run on Unix-like operating systems; Python (SciPy Python library used for scientific computing and technical computing); R environment for statistical computing and graphics; Eclipse integrated development environment; Redmine project management and issue tracking tool; X2Go remote desktop software.

CIE of RDWE delivers to researchers (to researcher's client device – laptop, desktop, mobile or tablet) using the extended *Platform-as-a-Service* service delivery model via X2Go remote desktop software and ssh cryptographic network protocol (picture).

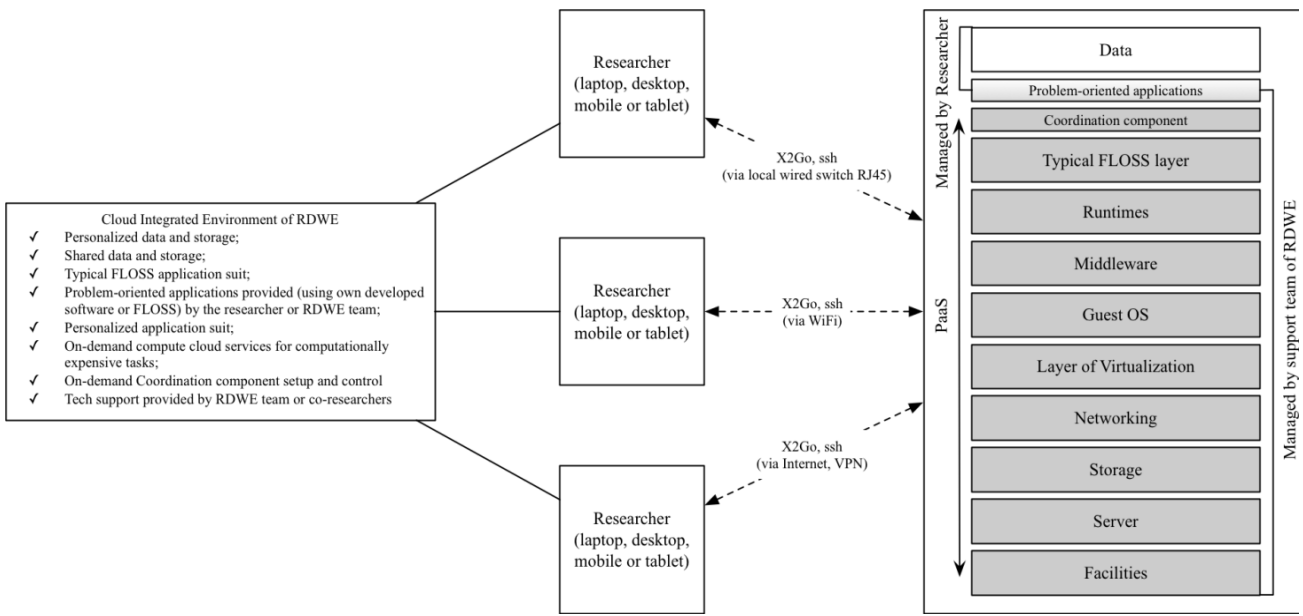

Picture. Cloud-integrated Environment of RDWE delivery model

To take all features of CIE, the researcher's client device (laptop or desktop) must run latest stable release of X2Go remote desktop software and comply with the following system requirements.

X2Go Client is part of Ubuntu 12.04 & later, Fedora 19 and later, Raspbian Wheezy & Jessie. X2Go Client is currently only released as a 32-bit x86 build. Both 32-bit x86 and 64-bit x86 versions of Windows are supported: Windows XP 32-bit SP3 (deprecated); Windows XP 64-bit SP2 (deprecated); Windows Vista SP2; Windows 7 SP1; Windows 8.1 (with "Update 1"); Windows 10 (1607).

Connecting to the CIE via ssh, the researcher's client device (laptop or desktop) must comply with the following system requirements:

- Windows XP 32-bit SP3,64-bit SP2; Vista SP2, 7 SP1, 8.1 (with "Update 1"), 10 (1607) with PuTTY terminal emulator installed;
- Linux;
- MacOS 10.9 Mavericks and higher were tested.

Software implementation of the RDWE class systems is carried out using software specific to the chosen domain area and may be different from the one above.

## Personal Research Information System – Research and Development Workstation Environment for ontology engineering and scientific creativity support

The main feature of the RDWE class systems is the problem orientation to an arbitrary domain area. As part of the research and development work of the Glushkov Institute of Cybernetics of National Academy of Sciences of Ukraine (Department of Microprocessor Technology) has developed and implemented the software system in its class. It was called – *Personal Research Information System* (PRIS) [33] – the RDWE class system for supporting research in the field of ontology engineering (the automated building of applied ontology in an arbitrary domain area as a main feature), scientific and technical creativity (the automated preparation of application documents for patenting inventions in Ukraine as a main feature). In accordance with the fundamental information model of the RDWE systems, the PRIS information model was developed.

At the actual stage of PRIS system development the $AWS$ set consists of the following $aws$ (natural language processing (NLP) functions are available for the Ukrainian language):

$aws_1$ – RESTful web service for *converting PDF files to plain/text*. Available via GitHub repository [34].

$aws_2$ – RESTful web service for *converting DOC/DOCX files to plain/text*. Available via GitHub repository [35].

$aws_3$ – RESTful web service for *language identification*.

$aws_4$ – RESTful web service for automatic *plain/text summarization*. Available via GitHub repository [36].

$aws_5$ – RESTful web service for *converting plain/text from UTF-8 to WIN-1251* enc. Available via GitHub repository [37].

$aws_6$ – RESTful web service for automatic *detection of title, author's names, and page numbers for PDF files*.

$aws_7$ – RESTful web service for automatic *keyword plain/text detection*.

$aws_8$ – RESTful web service for automatic *sentence segmentation*.

$aws_9$ – RESTful web service for automatic *word segmentation with lemmatization and Part-of-speech tagging*.

$aws_{10}$ – RESTful web service for automatic *word segmentation*.

$aws_{11}$ – RESTful web service for *Compositional Language Pre-processing* (CLP): term segmentation (dividing a string of written language into multiple-word and one-word terms) with lemmatization and Part-of-speech tagging.

$aws_{12}$ – RESTful web service for automatic *"stop words" filtering out*. These are words that do not bear the semantic load.

$aws_{13}$ – RESTful web service for automatic *word lemmatization*.

$aws_{14}$ – RESTful web service for *syntactic parser SyntaxNet: Neural Models of Syntax*: an open-source neural network framework implemented in TensorFlow that provides a foundation for Natural Language Understanding (NLU) systems [38 – 40].

$aws_{15}$ – RESTful web service for searching scientific publications in external bibliographic databases (an *intelligent agent for the automated search of scientific publications* [41]). Google Scholar search implemented via scholar.py [42] library.

$aws_{16}$ – RESTful web service for *text indexing and annotating* (for full-text search capability).

$aws_{17}$ – *MongoDB as a service* (storing the originals of text documents and texts, ontological structures, ontologies, processed texts in the form of JSON-documents, neural language vector models (NLVM) – word embedding models [43 – 44]).

$aws_{18}$ – *"Graph Editor"* – web service (represented as Single Page Application – SPA) for text's ontological representation, for manipulating ontologies and ontological structures. "Graph Editor" is a part of "TODOS" [45] – IT-platform formation transdisciplinary information environment.

$aws_{19}$ – *"CONFOR"* [46] – web service (represented as SPA) for intelligent data analysis. The main functions of "CONFOR" are: revealing the regularities that characterize classes of objects which are represented as sets of attribute values; using the revealed regularities for classification, diagnostics, and prediction.

$aws_{20}$ – *"KONSPEKT"* [47] – web service (represented as SPA) for syntactic and semantic analysis of natural language texts.

$aws_{21}$ – RESTful web service for processing NLVM of large corpora or open data collections (ODC): implemented via custom gensim [48] as a service server application. It is possible to [49]: calculate semantic similarity between pair of terms (including multiple-word terms, one-word terms, words) within the chosen NLVM; compute a list of nearest semantic associates for terms (including multiple-word terms, one-word terms, words) within the chosen NLVM; find the center of lexical cluster for a set of terms (including multiple-word terms, one-word terms, words) within the chosen NLVM; calculate semantic similarity between two sets of terms (including multiple-word terms, one-word terms, words) within the chosen NLVM.

$aws_{22}$ – "Personal ontological knowledge base for researcher's publications" web service [50] (represented as SPA).

$aws_{23}$ – Web service (represented as SPA) for creating and filling out templates that will allow to generate an incoming flow of documents coming from applicant of invention for industrial property [33].

$aws_{24}$ – additional external FLOSS (web services, libraries, utilities, etc.);

*The F set.* Functional filling-up of the PRIS is represented by the following set of synthesized from the set of AWS functions:

$$F = AWS:\{C_1, C_2, C_3, C_4, C_5, C_6, C_7\},$$

where:

$C_1$ – ontology engineering. The automated building of applied ontology in an arbitrary domain area (or knowledge domain). The fundamental foundations of the implemented technology are described in the works [50 – 55]:

$$C_1 = \{aws_1 \ldots aws_3, aws_5, aws_6, aws_8, aws_9, aws_{11}, aws_{12}, aws_{14}, aws_{16} \ldots aws_{20}\};$$

$C_2$ – the research design of scientific publications in an arbitrary domain area (or knowledge domain) [33, 50]:

$$C_2 = \{aws_1 \ldots aws_9, aws_{11} \ldots aws_{13}, aws_{15} \ldots aws_{18}, aws_{22}, aws_{24}\};$$

$C_3$ – personalized ontological processing of scientific publications [33, 50]:

$$C_3 = \{aws_1 \ldots aws_9, aws_{11} \ldots aws_{13}, aws_{15} \ldots aws_{18}, aws_{22}\};$$

$C_4$ – the automated CLP for training and querying NLVM for large corpora or ODC [33]:

$$C_4 = \{aws_1 \ldots aws_3, aws_5, aws_8 \ldots aws_{12}, aws_{20}, aws_{24}\};$$

$C_5$ – the automated distributive and semantic analysis of large corpora or ODC [33]:

$$C_5 = \{aws_{21}, aws_{24}\};$$

$C_6$ – the automatic syntactic parsing:

$$C_6 = \{aws_1 \ldots aws_3, aws_8, aws_{14}, aws_{20}, aws_{24}\};$$

$C_7$ – the automated preparation of application documents for patenting inventions in Ukraine. The fundamental foundations of the implemented technology are described in [33, 56]:

$$C_7 = \{aws_1 \ldots aws_3, aws_{21}, aws_{23}, aws_{24}\}$$

## Conclusion

The development of modern technologies increasingly covers the field of intellectual activity and, especially, in the field of scientific research and development. The existing Current Research Information Systems oriented on the following main types of services: access and reuse of scientific and academic information, methodologies, and technologies; information search; targeted dissemination of information; messaging services; bridging of horizontal and vertical relations between scientists; backup data storage and archival information.

We propose the new class of Current Research Information Systems and related intelligent information technologies. This class supports the main stages of the scientific research and development lifecycle, starting with the semantic analysis of the information of arbitrary domain area and ending with the formation of constructive features of innovative proposals. It was called – Research and Development Workstation Environment – the comprehensive problem-oriented information systems for scientific research and development support. A distinctive feature of such systems is the possibility of their problematic orientation to various types of scientific activities by combining on a variety of functional services and adding new ones within the Cloud-integrated Environment. Taking into account the objective factor of Ukrainian science state, we define the general principles of the RDWE design: modular design; microservices-based architecture; cross-platform software (also multi-platform software or platform-independent software); JSON/XML data model and REST architectural style (RESTful web services); using free and open-source software; cloud computing; data storage based on the abstraction of data repository; data repositories synchronization.

The Research and Development Workstation Environment class system's generalized information model is represented in the article as a three-tuple composite web service that include: a set of atomic web services, each of them can be designed and developed as a microservice or a desktop application, that allows them to be used as an independent software separately; a set of functions, the functional filling-up of the Research and Development Workstation Environment; a subset of atomic web services that are required to implement function of composite web service. In accordance with the fundamental information model of the Research and Development Workstation Environment systems, the Personal Research Information System information model was developed.

The main results of our work are focused on enhancing the effectiveness of the scientist's research and development lifecycle in the arbitrary domain area. In the future, it would be interesting to apply to the educational process the modern paradigm of Service-Oriented Learning implemented as Personal Learning Environments, Virtual Learning Environments, Learning Management Systems, Education-as-a-Service models, and the new class of E-learning systems – *Cloud Learning Environments* – using RDWE systems as a E-learning/education-oriented environment.

## References

1. Houssos N. 2011. CRIS for research information management.
2. ERGO – European Research Gateways Online 2018, viewed 07 February 2018, <https://cordis.europa.eu/news/rcn/8259_en.html>.
3. euroCRIS – The International Organization for Research Information 2018, viewed 07 February 2018, <https://www.eurocris.org>.
4. Main features of CERIF: The Common European Research Information Format 2018, viewed 07 February 2018, <https://www.eurocris.org/cerif/main-features-cerif>.
5. Modern scientific information systems. Prospects of use 2001, viewed 07 February 2018, <http://derpi.tuwien.ac.at/~andrei/CRIS_DOC.htm>.
6. Tan W. and Zhou M. 2013. Business and Scientific Workflows: A Web Service-Oriented Approach (Vol. 5). John Wiley & Sons.

7. Semenov M.A. 2017. Service support of e-learning at the replaced university. Open educational e-environment of modern university. N 3. P. 295–302.
8. Lindgren N. and Rautamki A. 2000. Managing strategic aspects of research. Proceedings CRIS-2000, Helsinki, ftp://ftp.cordis.lu/pub/cris2000/docs/rautamdki_fulltext. pdf.
9. Dew P., Leigh C., Whyte B. 2000. ADVISER II: Theory and practice of finding and presenting RTD results. Proceedings CRIS-2000, Helsinki, ftp://ftp.cordis.lu/pub/cris2000/docs/dewBfulltext.pdf.
10. CORDIS – Community Research and Development Information Service 2018, viewed 07 February 2018, <https://cordis.europa.eu/home_en.html>.
11. ResearchGate 2018, viewed 07 February 2018, <https://www.researchgate.net/>.
12. Helge P. 2013, 'A quick glance at business models of academic social networking services', *Hybrid Publishing Lab Notepad,* weblog post, 12 January, viewed 07 February 2018, <http://hybridpublishing.org/2013/01/a-quick-glance-at-business-models-of-academic-social-networking-services>.
13. Academia.edu – Share research 2018, viewed 07 February 2018, <https://www.academia.edu/>.
14. Mendeley – Reference Management Software & Researcher Network 2018, viewed 07 February 2018, <https://www.mendeley.com/>.
15. Van den Eynden V., Corti L., Woollard M., Bishop L. and Horton L. 2011. Managing and Sharing Data: a best practice guide for researchers.
16. Kothari C.R. 2004. Research methodology: Methods and techniques. New Age International.
17. Prokudin D. 2015. Design and Implementation of an Integrated Information System to Support Scientific Research. arXiv preprint arXiv:1504.04800.
18. ORCID – Connecting Research and Researchers 2018, viewed 07 February 2018, <https://orcid.org/>.
19. ResearcherID 2018, viewed 07 February 2018, <http://www.researcherid.com/>.
20. OpenID Foundation website 2018, viewed 07 February 2018, <https://openid.net/>.
21. Vestdam T. 2013. The future of CRIS systems – an interplay with VIVO.
22. Wiederhold L. 2000. Cooperative Structures for the Collection of Internet Resources on and from the Middle East. http://www.bibliothek.uni-halle.de/text/vortraege/venedig.htm.
23. Skorokhodov VA, Khudyakova I.M. Automated workplace manager: tutorial. Kyiv, 2008. 416 p.
24. Nabielsky J. and Skelton A.P. 1981. Virtual Terminal management model.
25. What are microservices? 2018, viewed 07 February 2018, <http://microservices.io/>.
26. Dvorkin E. 2014, 'Seven micro-services architecture advantages', *Art of Software Engineering,* weblog post, 3 June, viewed 07 February 2018, <http://eugenedvorkin.com/seven-micro-services-architecture-advantages/>.
27. Node.js 2018, *JavaScript runtime built on Chrome's V8 JavaScript engine*, viewed 07 February 2018, <https://nodejs.org>.
28. Fielding R.T. and Taylor R.N. 2000. Architectural styles and the design of network-based software architectures (Vol. 7). Doctoral dissertation: University of California, Irvine.
29. Etzkorn L.H. 2017. Introduction to Middleware: Web Services, Object Components, and Cloud Computing. CRC Press.
30. Bhowmik S. 2017. Cloud Computing. Cambridge University Press.
31. National Institute of Standards and Technology 2018, viewed 07 February 2018, <https://www.nist.gov/>.
32. Shkarupylo V., Kudermetov R., Paromova T. 2012. Conceptual model of automated composite web services synthesis process.
33. Palagin O.V., Velychko V.Yu., Malakhov K.S. and Shchurov O.S. 2017. Personal research information system. About developing the methods for searching patent analogs of invention. Computer means, networks and systems. N 16. P. 5–13.
34. Atomic Web Service for converting PDF files to plain/text 2018, viewed 07 February 2018, <https://github.com/malakhovks/pdf-extract-api>.
35. Atomic Web Service for converting DOC/DOCX files to plain/text 2018, viewed 07 February 2018, <https://github.com/malakhovks/doc-docx-extract-api>.
36. Atomic Web Service for automatic text summarization 2018, viewed 07 February 2018, <https://github.com/malakhovks/text-summarization-api>.
37. Atomic Web Service for converting text from UTF-8 to WIN-1251 2018, viewed 07 February 2018, <https://github.com/malakhovks/utf8-to-win1251-api>.
38. Petrov S. 2016. 'Announcing SyntaxNet: The World's Most Accurate Parser Goes Open Source', *Google Research Blog,* weblog post, 12 May, viewed 07 February 2018, <https://research.googleblog.com/2016/05/announcing-syntaxnet-worlds-most.html>.
39. SyntaxNet: Neural Models of Syntax 2018, viewed 07 February 2018, <https://github.com/tensorflow/models/tree/master/research/syntaxnet>.
40. TensorFlow – An open-source machine learning framework for everyone 2018, viewed 07 February 2018, <https://www.tensorflow.org/>.
41. Velychko V.Yu., Malakhov K.S., Shchurov O.S. 2017. Information model of an intelligent agent-aided search of scientific papers. *Computational Intelligence (Results, Problems and perspectives),* P. 105–107.
42. ckreibich/scholar.py: A parser for Google Scholar, written in Python 2018, viewed 07 February 2018, <https://github.com/ckreibich/scholar.py>.
43. Mikolov T., Chen K., Corrado G. and Dean J. 2013. Efficient estimation of word representations in vector space. arXiv preprint arXiv:1301.3781.
44. Bojanowski P., Grave E., Joulin A. and Mikolov T. 2016. Enriching word vectors with subword information. arXiv preprint arXiv:1607.04606.
45. Velychko V.Yu., Malakhov K.S., Semenkov V.V., Strizhak A. E. 2014. Integrated Tools for Engineering Ontologies. *Information Models and Analyses.* N 4. P. 336–361.
46. Gladun V., Velychko V., Ivaskiv Y. 2008. Selfstructurized Systems. *International Journal "Information Theories & Applications"*, Vol. 15. N 1. P. 5–13.
47. Andrushhenko T.I., Velychko V.Yu., Hal"chenko S.A., Hloba L.S., Hulyayev K.D., Klimova E.Ya., Komova O.B., Lisovyj O.V., Popova M.A., Pryxod¬nyuk V.V., Stryzhak O.Ye., Stus D.M., 2013. Metodyky napysannja naukovyx robit na osnovi ontolohichnoho analizu tekstiv: metodychnyj posibnyk K.: TOV «SITIPRINT», 124 p. (In Ukrainian).
48. RaRe-Technologies/gensim: Topic Modelling for Humans 2018, viewed 07 February 2018, <https://github.com/RaRe-Technologies/gensim>.
49. Kutuzov A. and Kuzmenko E. 2016, April. WebVectors: a toolkit for building web interfaces for vector semantic models. In International Conference on Analysis of Images, Social Networks and Texts (P. 155–161). Springer, Cham.
50. Palagin O.V., Velychko V.Yu., Malakhov K.S. and Shchurov O.S. 2017. Design and software implementation of subsystems for creating and using the ontological base of a research scientist. *Problems in programming.* N 2. P. 72–78.
51. Palagin A.V., Petrenko N.G., Malakhov K.S. 2011. Technique for designing a domain ontology. *Computer means, networks and systems*, N 10. P. 5–12.
52. Palagin O.V., Petrenko M.G. and Kryvyi S.L. 2012. Ontolohichni metody ta zasoby obrobky predmetnykh znan. Publishing center of V. Dahl East Ukrainian National University.
53. Palagin A.V., Petrenko N.G., Velychko V.Y. and Malakhov K.S. 2018. The problem of the development ontology-driven architecture of intellectual software systems. arXiv preprint arXiv:1802.06767.
54. Palagin A.V., Petrenko N.G., Velichko V.Yu. and Malakhov K.S. 2014. Development of formal models, algorithms, procedures, engineering and functioning of the software system "Instrumental complex for ontological engineering purpose". *Problems in programming*, N 2–3. P. 221–232.

55. Palagin A.V., Petrenko N.G., Velichko V.Yu., Malakhov K.S. and Tikhonov Yu.L. 2012. To the problem of "The Instrumental complex for ontological engineering purpose" software system design. *Problems in programming*, N 2–3. P. 289–298.
56. Palagin, O., Kurhaiev, O., Petrenko, M., Semotiuk, M., Velychko, V., Marchenko, V., Malakhov, K., Yakovlev, Yu., Kurzantseva, L., Sosnenko, K., Samoliuk, T., Hryhoriev. S., 2017. Rozrobka elementiv tekhnolohii informatsiino-ontolohichnoi pidtrymky naukovo-tekhnichnoi tvorchosti. *Project report 205.38.17*.

## Література

1. Houssos N. CRIS for research information management. URL: http://hdl.handle.net/11366/303 (Last accessed: 07.02.2018).
2. ERGO – European Research Gateways Online. URL: https://cordis.europa.eu/news/rcn/8259_en.html (Last accessed: 07.02.2018).
3. euroCRIS – The International Organization for Research Information. URL: https://www.eurocris.org (Last accessed: 07.02.2018).
4. Main features of CERIF: The Common European Research Information Format. URL: https://www.eurocris.org/cerif/main-features-cerif (Last accessed: 07.02.2018).
5. Лопатенко А.С. Современные научные информационные системы. Перспективы использования. URL: http://derpi.tuwien.ac.at/~andrei/CRIS_DOC.htm (дата звернення: 07.02.2018).
6. Wei Tan, MengChu Zhou. Business and Scientific Workflows: A Web Service-Oriented Approach. New Jersey, 2013. 264 p.
7. Семенов М.А. Сервісний супровід дистанційного навчання в переміщеному університеті. *Відкрите освітнє е-середовище сучасного університету*. 2017. № 3. С. 295—302.
8. Lindgren N., Rautamki A. Managing strategic aspects of research. *Proceedings CRIS-2000, Helsinki*. URL: https://pdfs.semanticscholar.org/ca88/763a60bf5adfa32aaa67bc284a842f4c2887.pdf (Last accessed: 07.02.2018).
9. Dew P., Leigh C., Whyte B. ADVISER II:Theory and practice of finding and presenting RTD results. *Proceedings CRIS-2000, Helsinki*. URL: ftp://ftp.cordis.lu/pub/cris2000/docs/dewBfulltext.pdf (Last accessed: 07.02.2018).
10. CORDIS – Community Research and Development Information Service. URL: https://cordis.europa.eu/home_en.html (Last accessed: 07.02.2018).
11. ResearchGate. URL: https://www.researchgate.net/ (Last accessed: 07.02.2018).
12. Helge P. A quick glance at business models of academic social networking services. *Hybrid Publishing Lab Notepad*. URL: http://hybridpublishing.org/2013/01/a-quick-glance-at-business-models-of-academic-social-networking-services/ (Last accessed: 07.02.2018).
13. Academia.edu – Share research. URL: https://www.academia.edu/ (Last accessed: 07.02.2018).
14. Mendeley – Reference Management Software & Researcher Network. URL: https://www.mendeley.com/ (Last accessed: 07.02.2018).
15. Van den Eynden V., Corti L., Woollard M., Bishop L., Horton L. Managing and sharing data: best practice for researchers. UK, 2011. 40 p.
16. Kothari C.R., Guarav Garg. Research Methodology: Methods and Techniques 3rd edition. New Delhi, 2014. 449 p.
17. Прокудин Д.Е. Проектирование и реализация комплексной информационной системы поддержки научных исследований. *Технологии информационного общества в науке, образовании и культуре: сборник научных статей*. Материалы научной конференции "Интернет и современное общество" IMS-2014 (Санкт-Петербург 19—20 ноября 2014). Санкт-Петербург, 2014. С. 31—36.
18. ORCID – Connecting Research and Researchers. URL: https://orcid.org/ (Last accessed: 07.02.2018).
19. ResearcherID. URL: http://www.researcherid.com/ (Last accessed: 07.02.2018).
20. OpenID Foundation website. URL: https://openid.net/ (Last accessed: 07.02.2018).
21. Vestdam T. The future of CRIS systems - an interplay with VIVO. *4th Annual VIVO Conference* (St Louis, Aug 14-16, 2013). URL: http://hdl.handle.net/11366/334 (Last accessed: 07.02.2018).
22. Wiederhold L. Cooperative Structures for the Collection of Internet Resources on and from the Middle East. URL: http://www.bibliothek.uni-halle.de/text/vortraege/venedig.htm (Last accessed: 07.02.2018).
23. Скороходов В.А., Худякова І.М. Автоматизоване робоче місце менеджера: навчальний посібник. Київ, 2008. 416 с.
24. Nabielsky J., Skelton A.P. A Virtual Terminal Management Model. URL: https://www.rfc-editor.org/pdfrfc/rfc782.txt.pdf (Last accessed: 07.02.2018).
25. What are microservices? URL: http://microservices.io/ (Last accessed: 07.02.2018).
26. Dvorkin E. Seven micro-services architecture advantages. *Art of Software Engineering*. URL: http://eugenedvorkin.com/seven-micro-services-architecture-advantages/ (Last accessed: 07.02.2018).
27. Node.js – JavaScript runtime built on Chrome's V8 JavaScript engine. URL: https://nodejs.org (Last accessed: 07.02.2018).
28. Fielding R.T., Taylor R.N. Architectural styles and the design of network-based software architectures. *Doctoral dissertation: University of California, Irvine, 2000 Vol. 7*. URL: https://www.ics.uci.edu/~fielding/pubs/dissertation/fielding_dissertation.pdf (Last accessed: 07.02.2018).
29. Etzkorn L.H. Introduction to Middleware: Web Services, Object Components, and Cloud Computing. CRC Press, 2017. 662 p.
30. Bhowmik S. Cloud Computing. Cambridge University Press, 2017. 462 p.
31. National Institute of Standards and Technology. URL: https://www.nist.gov/ (Last accessed: 07.02.2018).
32. Шкарупило В. В., Кудерметов Р. К., Паромова, Т. А. Концептуальная модель процесса автоматизированного синтеза композитных веб-сервисов. *Наукові праці Донецького національного технічного університету*. Сер.: Інформатика, кібернетика та обчислювальна техніка. 2010. Випуск 15. С. 231—238.
33. Палагін О. В., Величко В. Ю., Малахов К. С., Щуров О. С. Автоматизоване робоче місце наукового дослідника. До питання розробки методів пошуку аналогів патентної документації винаходу. *Комп'ютерні засоби, мережі та системи*. 2017. № 16. С. 5—13.
34. Atomic Web Service for converting PDF files to plain/text. URL: https://github.com/malakhovks/pdf-extract-api (Last accessed: 07.02.2018).
35. Atomic Web Service for converting DOC/DOCX files to plain/text. URL: https://github.com/malakhovks/doc-docx-extract-api (Last accessed: 07.02.2018).
36. Atomic Web Service for automatic text summarization. URL: https://github.com/malakhovks/text-summarization-api (Last accessed: 07.02.2018).
37. Atomic Web Service for converting text from UTF-8 to WIN-1251. URL: https://github.com/malakhovks/utf8-to-win1251-api (Last accessed: 07.02.2018).
38. Petrov S. Announcing SyntaxNet: The World's Most Accurate Parser Goes Open Source. *Google Research Blog*. URL: https://research.googleblog.com/2016/05/announcing-syntaxnet-worlds-most.html (Last accessed: 07.02.2018).
39. SyntaxNet: Neural Models of Syntax. URL: https://github.com/tensorflow/models/tree/master/research/syntaxnet (Last accessed: 07.02.2018).
40. TensorFlow – An open-source machine learning framework for everyone. URL: https://www.tensorflow.org/ (Last accessed: 07.02.2018).
41. Величко В.Ю., Малахов К.С., Щуров О.С. Информационная модель интеллектуального агента автоматизированного поиска научных работ. *Обчислювальний інтелект (Результати, Проблеми, Перспективи)*: Матеріали 4-ої міжнар. наук-практ. конф. (Київ, 16–18 травня 2017). Київ, 2010. С. 105–107.
42. ckreibich/scholar.py: A parser for Google Scholar, written in Python. URL: https://github.com/ckreibich/scholar.py (Last accessed: 07.02.2018).
43. Mikolov T., Chen K., Corrado G., Dean J. Efficient Estimation of Word Representations in Vector Space. *arXiv preprint arXiv:1301.3781*. URL: https://arxiv.org/abs/1301.3781 (Last accessed: 07.02.2018).

44. Bojanowski P., Grave E., Joulin A., Mikolov, T. Enriching word vectors with subword information. *arXiv preprint arXiv:1607.04606*. URL: https://arxiv.org/abs/1607.04606 (Last accessed: 07.02.2018).
45. Величко В. Ю., Малахов К. С., Семенков В. В., Стрижак, А. Е. Комплексные инструментальные средства инженерии онтологий. *Information Models and Analyses.* 2014. № 4. С. 336–361.
46. Gladun V., Velychko V., Ivaskiv Y. Selfstructurized Systems. // International Journal "Information Theories & Applications". FOI ITHEA, Sofia. 2008. Vol. 15. N 1. P. 5–13.
47. Андрущенко Т.І., Величко В.Ю., Гальченко С.А., Глоба Л.С., Гуляєв К.Д., Клімова Е.Я., Комова О.Б., Лісовий О.В., Попова М.А., Приходнюк В.В., Стрижак О.Є., Стус Д.М. Методики написання наукових робіт на основі онтологічного аналізу текстів: методичний посібник К. : ТОВ «СІТІПРІНТ», 2013. 124 с.
48. RaRe-Technologies/gensim: Topic Modelling for Humans. URL: https://github.com/RaRe-Technologies/gensim (Last accessed: 07.02.2018).
49. Kutuzov A., Kuzmenko E. (2017) WebVectors: A Toolkit for Building Web Interfaces for Vector Semantic Models. In: Ignatov D. et al. (eds) Analysis of Images, Social Networks and Texts. AIST 2016. Communications in Computer and Information Science, vol 661. Springer, Cham.
50. Палагін О. В., Величко В. Ю., Малахов К. С., Щуров О. С. Проектування та програмна реалізація підсистеми створення та використання онтологічної бази знань публікацій наукового дослідника. *Проблеми програмування.* 2017. № 2. С. 72–78.
51. Палагін А. В., Петренко Н. Г., Малахов, К. С. Методика проектирования онтологии предметной области. *Комп'ютерні засоби, мережі та системи.* 2011. № 10. С. 5–12.
52. Палагін О. В., Петренко М. Г., Кривий, С. Л. Онтологические методы и средства обработки предметных знаний. Луганськ, 2012. 323 с.
53. Палагін А. В., Петренко Н. Г., Величко В. Ю., Малахов К. С. К вопросу разработки онтолого-управляемой архитектуры интеллектуальной программной системы. *Вісник Східноукраїнського національного університету імені Володимира Даля.* 2011. № 3. С. 179–184.
54. Палагін А.В., Петренко Н.Г., Величко В.Ю., Малахов, К.С. Развитие формальных моделей, алгоритмов, процедур, разработки и функционирования программной системы "Инструментальный комплекс онтологического назначения". *Проблеми програмування.* 2014. № 2–3. С. 221–232.
55. Палагін А.В., Петренко Н.Г., Величко В.Ю., Малахов К.С., Тихонов, Ю.Л. К вопросу разработки инструментального комплекса онтологического назначения. *Проблеми програмування.* 2012. № 2–3. С. 289–298.

*About the authors:*

*Oleksandr Palagin,*
Doctor of Sciences, Academician of National Academy of Sciences of Ukraine,
Deputy director of Glushkov Institute of Cybernetics,
head of department 205 at Glushkov Institute of Cybernetics,
290 Ukrainian publications,
45 International publications,
H-index: Google Scholar – 15,
Scopus – 3.
http://orcid.org/0000-0003-3223-1391.

*Vitalii Velychko,*
PhD, assistant professor, Senior researcher,
73 Ukrainian publications,
25 International publications.
H-index: Google Scholar – 7,
Scopus – 1.
http://orcid.org/0000-0002-7155-9202.

*Kyrylo Malakhov,*
Junior Research Fellow,
32 Ukrainian publications,
2 International publications.
H-index: Google Scholar – 4.
http://orcid.org/0000-0003-3223-9844.

*Oleksandr Shchurov,*
1 category software engineer,
6 Ukrainian publications.
H-index: Google Scholar – 1.
http://orcid.org/0000-0002-0449-1295.

*Affiliation:*

Glushkov Institute of Cybernetics of National Academy of Sciences of Ukraine,
40 Glushkov ave., Kyiv, Ukraine, 03187.
Tel.: (+38) (044) 526 3348.
Email: palagin_a@ukr.net,
    aduisukr@gmail.com